\documentclass[twocolumn,showpacs,amssymb,nobibnotes,aps,floats,psfig,prb]{revtex4}
\usepackage{hyperref}
\usepackage{amsbsy}
\usepackage{amsmath}
\usepackage{graphicx,epsfig}
\usepackage{epsfig}
\usepackage{float}
\def \be{\begin{equation}} 
\def \ee{\end{equation}} 
\def \bea{\begin{eqnarray}} 
\def \eea{\end{eqnarray}}
\begin{document}
\title{Quantum phase transition in Bose-Holstein model in two dimensions}
\author{Sanjoy Datta  }
\email{sanjoy.datta@saha.ac.in}
\author{Sudhakar Yarlagadda}
\email{y.sudhakar@saha.ac.in}
\affiliation{ Saha Institute of Nuclear Physics, 1/AF-Bidhannagar,
Kolkata-64, India}
\date{\today}

\begin{abstract}
We  derive an effective d-dimensional Hamiltonian for a 
system of hard-core-bosons coupled to optical phonons in a lattice. 
Away from half-filling, we show that the presence of next-nearest-neighbor
hopping in the effective Hamiltonian
leads to a superfluid-to-supersolid transition at intermediate
boson-phonon (b-p) couplings, while at strong-couplings the system phase separates.
However, at half-filling and at a critical b-p coupling
(as in the xxz-model), 
 the system undergoes a superfluid-to-charge-density-wave transition
without any signature of supersolidity.
Our analyses is based on extensive calculations of the structure factor, the
superfluid fraction, the Bose-Einstein condensate fraction, and the system energy
at various fillings.
We present a phase diagram for this system and compare it to that of the xxz-model. 
We also demonstrate explicitly that the 
next-nearest-neighbor hopping (in the absence of nearest-neighbor hopping)
 in the effective Hamiltonian leads only to a single
transition -- a  first-order superfluid-to-supersolid transition.
\end{abstract}
\maketitle

\nopagebreak

\section{INTRODUCTION}\label{intro}
The successful mimicking of an actual lattice using optical standing waves
marks one of the most significant scientific advances of this decade
\cite{grein1}. 
The biggest advantage of this kind of 
an optical lattice is that the ratio of
the
kinetic energy and the interaction energy of
the particles 
can be controlled at will. 
This has led to a flurry of activities among atomic 
and condensed matter physicists across the world. 
The excitement among the condensed matter physicists stems from the fact that
 it not only gives a testing ground for some of the most intriguing phenomena
 of nature predicted earlier, but it also paves the way for the
 discovery of new physical phenomena.
For example, soon after the creation of a two
 dimensional (2D) optical lattice, it has been used to experimentally verify
\cite{grein2} the predicted transition \cite{M.P.A.Fisher}
 from a superfluid state
to a Mott insulating state of a
 bosonic system.
Another possibility is the verification
of the theoretically predicted supersolidity due to vacancies
 \cite{Andreev,Chester,Leggett}.
 A signature of supersolidity 
is
the simultaneous
 presence of both diagonal long range order (DLRO) and off diagonal long range
 order (ODLRO) \cite{penrose,Onsager}. 
 There have not been many studies of this interesting
 phase of
 matter until it was recently observed
in helium-4 \cite{KimChan-Nature,KimChan-Science}.
 This discovery led theorists to study bosonic models 
 in different kinds of lattice structures
\cite{Melko05,Damle,Wessel}.
 and with various types 
of interactions among these particles 
\cite{Sengupta}.

There has been very little attention  given to a system of bosons interacting 
with phonons. Recently Pupillo {\em et al.} \cite{pupillo} have
 studied such a possibility where the bosons could be coupled to the
 acoustic phonons generated by polar molecules trapped to form a lattice. 
In this paper, we have considered a Bose-Holstein model comprising
of hard-core-bosons (hcb) coupled to optical
phonons generated by the vibrations of the underlying lattice.
An example of such hcb is a collection of tightly-bound Cooper pairs
originating from
electronic polarization processes \cite{varma,tvr}.
Additionally,
 strong-coupling between
electrons and intermolecular-phonons also
produces hcb; when
such hcb couple to intra-molecular phonons, the system can be studied by
 a Bose-Holstein model
\cite{ramys}.
 Starting with a minimalistic model, involving momentum
independent b-p coupling, we have 
derived an effective d-dimensional Hamiltonian for hcb 
by using a transparent non-perturbative technique.
The region of validity of our effective Hamiltonian is governed by the
small parameter ratio of the adiabaticity $t/\omega_0$ and the
b-p coupling $g$.
 The most interesting feature of this 
effective Hamiltonian is that, besides a nearest-neighbor (NN) hopping,
 it consists of next-nearest-neighbor (NNN)
hopping and NN repulsion. 
Our approach gives a microscopic justification for the origin
of these important additional terms.
We study our derived effective Hamiltonian
in 2D by using exact 
diagonalization technique.
For exact diagonalization, we have used a modified 
Lanczos algorithm \cite{gagli} on lattice clusters
with 
$4 \times 4$, $\sqrt{18} \times \sqrt{18}$, and  
$\sqrt{20} \times \sqrt{20}$ sites.
 We have shown that, except for the extreme anti-adiabatic limit,
 the hcb coupled with optical phonons can show supersolidity above a 
critical value of the b-p coupling strength. 

The paper has been arranged as follows.
In Sec. \ref{Effective_Hamil},
we have derived the effective Hamiltonian for a system of hcb 
coupled to optical phonons. We discuss briefly the basic difference of this
effective Hamiltonian with that for fermions\cite{sdadys}.
Next, 
 we apply mean field analysis to this Hamiltonian
in Sec. \ref{MFA}
 and obtain a
mean field 
phase diagram.
In Sec. \ref{DLRO}, we discuss in detail the DLRO
 by studying the structure factor. Here, we also  present  key
numerical results. In Sec. \ref{ODLRO},
 we discuss two important quantities --
the Bose condensate fraction and the superfluid fraction. Sec. \ref{maxwell}
deals with calculating the free energy of the system for different situations
 and parameter values. The curvature of the free-energy-versus-filling
curves is used in deciding whether
the system phase separates or not.
 Finally, in Sec. \ref{Results}, we present the results.

\section{\label{Effective_Hamil}Effective Hamiltonian}
\begin{figure}[]
\includegraphics*[width=0.8\linewidth]{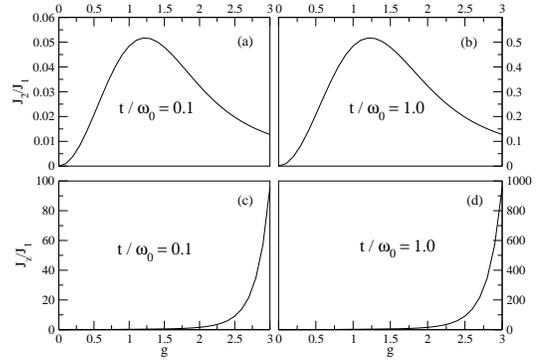}
\caption{Comparison of $J_2/J_1$ and $J_z/J_1$ at two different values of
 $t/\omega_0$ and various values of g.}
\label{constants}
\end{figure}

We start with a system of spinless hcb coupled with 
optical phonons on a square lattice. This system is described by a 
Bose-Holstein Hamiltonian 
\cite{holstein}
\be
H_{\textrm{hol}} = -t \sum_{j,\delta} b^{\dagger}_{j+\delta} b_j 
          + \omega_0 \sum_j a^{\dagger}_{j} a_j
          + g \omega_0 \sum_j n_j
 (a_j +a^{\dagger}_j) , 
\ee
where $\delta$ corresponds to nearest-neighbors,
 $n_j \equiv b^{\dagger}_{j} b_j $ with 
 $b_j$ being the destruction operator for hcb
 (and not of electrons as in the
 Holstein model), while (as in the Holstein case)
$a_j$ is the destruction operator for phonons,
and $\omega_0$ is the single vibrational
 frequency for simple harmonic oscillators. Then we perform 
the Lang-Firsov (LF) transformation \cite {lang} on this Hamiltonian which produces displaced
simple harmonic oscillators and dresses the hopping particles with phonons.
It is important to note that although we 
are dealing with particles different from fermions, 
we can still perform the same LF transformation.
This is because, under the
 LF transformation  given by $e^S H_{\textrm{hol}} e^{-S} $ with
$S= - g \sum_i n_i (a_i - a^{\dagger}_i)$,
$b_j$ and $a_j$ transform (like fermions  and phonons in the Holstein model) as follows:
\bea
\tilde b_j &\equiv& e^S b_j e^{-S} = b_j e^{-g(a_j-a^{\dagger}_j)} ,
 \nonumber \\
\tilde a_j &\equiv& e^S a_j e^{-S} = a_j - g n_j .
\label{trans}
\eea
 This is due to the unique (anti-) commutation properties of hcb
 given by 
\bea
[b_i,b_j]&=&[b_i,b^{\dagger}_j]= 0 , \textrm{ for } i \neq j , \nonumber\\
\{b_i,b^{\dagger}_i\}& = & 1 .
\label{commute}
\eea     
Next, we take the 
 unperturbed Hamiltonian to be given by
\cite{sdadys}
\be
H_0 = \omega_0 \sum_j a^{\dagger}_j a_j 
      - g^2 \omega_0 \sum_j b^{\dagger}_j b_j
      -J_1 \sum_j (b_j b_{j+\delta}+ {\rm H.c.}) , 
\ee 
and the perturbation to be
\be
H'=\sum_j H_j = -J_1 \sum_j(b^{\dagger}_j b_{j+\delta}
            \{\mathcal S^{{j}^\dagger}_+ \mathcal S^{j}_{-}-1\} + {\rm H.c.}),
\ee
where $\mathcal S^{j}_{\pm} = \textrm{exp}[\pm g(a_j - a_{j+\delta})],
J_1 = t \textrm{exp}(-g^2) $, and $g^2 \omega_0 $ is the polaronic binding energy.
Here, $H_0 + H'$ constitutes the LF transformed Bose-Holstein Hamiltonian.
 We then follow the same 
steps as in Ref. [\onlinecite{sdadys}] to get the following effective Hamiltonian 
in d-dimensions for our Bose-Holstein model
\bea
H_{eff} &=& -g^2 \omega_0 \sum_j n_j - 
         J_1 \sum_{j,\delta} b^{\dagger}_j b_{j+\delta} \nonumber \\
        & & - J_2 \sum_{j,\delta,\delta' \neq \delta}
             b^{\dagger}_{j+\delta'}b_{j+\delta} 
         - 0.5 J_z \sum_{j,\delta} n_j(1- n_{j+\delta}) , \nonumber \\ 
        & & 
\label{heff}
\eea  
where $J_z \equiv (J_1^2/\omega_0)[4 f_1(g)+2 f_2(g)]$ and 
$J_2 \equiv (J_1^2/\omega_0)f_1(g)$ with 
$f_1(g) \equiv \sum^{\infty}_{n=1} g^{2n}/(n!n)$ and
$f_2(g) \equiv \sum^{\infty}_{n=1}\sum^{\infty}_{m=1} g^{2(n+m)}/[n!m!(n+m)]. $
In Fig. \ref{constants}, we plot the ratios $J_2/J_1$ and $J_z/J_1$ for
various values of $g$ and adiabaticity parameter $t/\omega_0$. We note
that both 
 $(J_2/J_1)\times (\omega_0/t)$ and 
 $(J_z/J_1)\times (\omega_0/t)$ are functions of $g$ only. 

\begin{figure}[]
\includegraphics*[width=0.8\linewidth]{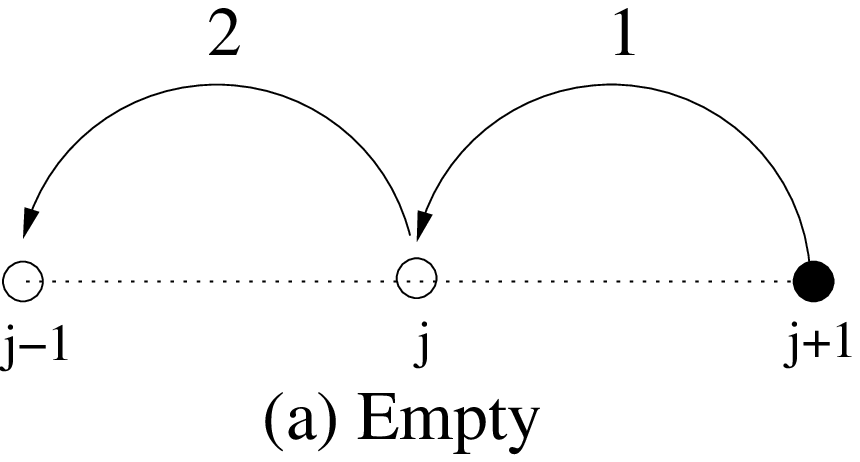}

\vspace{0.3cm}

\includegraphics*[width=0.8\linewidth]{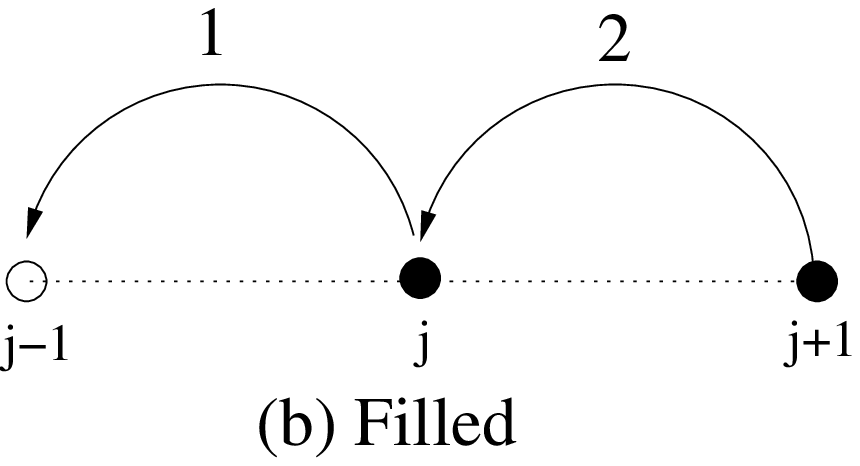}
\caption{Depicted processes describe the following terms:
(a) $c^{\dagger}_{j-1} c_j c^{\dagger}_j c_{j+1}$ when site $j$ is empty; and
(b) $c^{\dagger}_j c_{j+1} c^{\dagger}_{j-1} c_j $ for a filled site $j$}.
\label{hopp2}
\end{figure}

The effective Hamiltonian in Eq. (\ref{heff})
 is different from that for spinless fermions in Ref. [\onlinecite{sdadys}].
This is because the effective Hamiltonian for fermions contains 
an {\it extra correlated hopping term}
$J_2 \sum_{j,\delta,\delta' \neq \delta} 2 n_j
 c^{\dagger}_{j+\delta'}c_{j+\delta}$ (with $c_j$ being the destruction operator
for fermions) because the commutation relations for fermions are different
from those of hcb given in
 Eq. (\ref{commute}).
 To see the difference clearly, let us consider the simplest case of 
one-dimension (1D). After carrying out the second-order perturbation theory
for fermions (bosons),
we get in 1D
the term $c^{\dagger}_{j-1} c_j c^{\dagger}_j c_{j+1} $ 
($b^{\dagger}_{j-1} b_j b^{\dagger}_j b_{j+1}$) depicted by the process (a)
in Fig. \ref{hopp2}
 and the term 
$c^{\dagger}_j c_{j+1} c^{\dagger}_{j-1} c_j $ 
($b^{\dagger}_j b_{j+1} b^{\dagger}_{j-1} b_j$) depicted by the process (b)
 in Fig. \ref{hopp2}.
 For fermions, when
these two terms are added, one gets 
$c^{\dagger}_{j-1}(1-2 n_j) c_{j+1}$ whereas for bosons one gets only
$b^{\dagger}_{j-1} b_{j+1}$.
 These arguments can easily be extended to d-dimensions.
 In a previous work \cite{fradkin}, while performing a similar second-order perturbation theory,
the authors missed the 
 process depicted in Fig. \ref{hopp2}(b).
Here we would like to point out that, as mentioned in Ref. \onlinecite{sudhakar1},
the small parameter for
our perturbation theory is $t/(g \omega_0)$.
\section{\label{MFA}Mean field analysis}
In this section, we shall study the phase transitions 
dictated by the effective Hamiltonian of Eq. (\ref{heff}) 
by employing the mean filed analysis (MFA) of Robaszkiewicz 
{\em et al.} \cite{micnas}.
We first note that the hcb
 may be represented by spin one-half
operators. More precisely,
 with the transformations $ S^{+} = S^x +  i S^y =  b^{\dagger}$, 
$S^{-} = S^x -  i S^y =  b$, and 
$S^z + 0.5 = b^{\dagger}b$, the commutation relations of Eq. (\ref{commute}) are preserved. 
We can then write the Hamiltonian of Eq. (\ref{heff}) in the following form:
\bea
H_0 &=& -J_1 \sum_{j,\delta}^{} (S_{j}^{x}S_{j+\delta}^{x}
+S_{j}^{y}S_{j+\delta}^{y}) \nonumber \\
& & - J_2\sum_{j,\delta,\delta',\delta \neq \delta'}
(S_{j+\delta'}^{x}S_{j+\delta}^{x}+S_{j+\delta'}^{y}S_{j+\delta}^{y}) \nonumber \\
& & + 0.5 J_z \sum_{j,\delta} S_{j}^{z}S_{j+\delta}^{z}
- B \sum_{j} (2 S_{j}^{z}+1) , 
\label{ham_mfa}
\eea
with the constraint
\begin{equation}
\frac{1}{N}\sum_i \left< S^{z}_i \right> = \frac{1}{2}(2n-1) ,
\label{constraint}
\end{equation}
where $N$ is the number of sites in the lattice.
Here, $B=J_{z}/2+g^2\omega_0/2$ is the effective magnetic field and $n = \frac{1}{N}\sum_i \left< b^{\dagger}_i b_i \right>$ is the filling
fraction ($0\leq n \leq 1$).
 In the MFA, for a trial Hamiltonian $H_0$
the following identity holds:
\begin{equation}
 F \leq F_0 = -\frac{1}{\beta} \textrm{ln Tr}\left[\textrm{exp}
(-\beta H_0)\right] + 
\left< H - H_0 \right>_0 ,
\end{equation}
where $\beta = 1/k_B T$ and $\left< \dots \right>_0$ is the thermal average 
with respect to the trial Hamiltonian $H_0$ .
The trial Hamiltonian $H_0$ is chosen as
\begin{equation}
H_0 = -\sum_i \vec{\Lambda_i}.\vec{S_i} - B \sum_i 1 ,
\label{mfa_H_0}
\end{equation}
where the molecular fields $ \vec{\Lambda_i}$ are obtained variationally
by minimizing $F_0$. 
 After some standard calculation, we obtain
\bea
\Lambda^{x}_{i} = \Lambda^{y}_{i} &=&
2 \sum_{\delta} J_1 \left< S^{x}_{i+\delta} \right>_0 +
2 \sum_{\delta,\delta',\delta \neq \delta'} J_2 \left< S^{x}_{i+\delta - \delta'} \right>_0 \nonumber \\
 \Lambda^{z}_{i} &=&  2B - 
 \sum_{\delta} 
J_z \left< S^{z}_{i+\delta} \right>_0 .
\eea
The eigenenergies of Eq. (\ref{mfa_H_0}) are 
\begin{equation}
 \lambda = -B \pm \Delta_{\zeta},
\label{eigenenrgies}
\end{equation}
where
\begin{equation}
\Delta_{\zeta} = \sqrt{\left(\frac{\Lambda^{z}_{\zeta}}{2}\right)^2 +
\left(\frac{\Lambda^{x}_{\zeta}}{2}\right)^2}  .
\label{delta}
\end{equation}
Here $\zeta = a,b $ represents the two sub-lattices. The eigenfunctions
are given by
\bea
\psi_{\zeta}^{+} &=& \textrm{cos}(\frac{\theta_\zeta}{2})|\frac{1}{2}> +
            \textrm{sin}(\frac{\theta_\zeta}{2})|-\frac{1}{2}>  \nonumber \\
\psi_{\zeta}^{-} &=& -\textrm{sin}(\frac{\theta_\zeta}{2})|\frac{1}{2}> +
              \textrm{cos}(\frac{\theta_\zeta}{2})|-\frac{1}{2}>   ,
\label{eigenfunc}
\eea
where 
$\textrm{sin}\theta_\zeta = \frac{\Lambda^{x}_{\zeta}}{2\Delta_{\zeta}} 
\textrm{ and }
\textrm{cos}\theta_\zeta = \frac{\Lambda^{z}_{\zeta}}{2\Delta_{\zeta}}$.
At ${\rm T=0}$K, ground state expectation value of $S_x$ and $S_z$ are given by
\bea
\left< S^{x}_{\zeta} \right>_0 &=&\frac{\textrm{sin}\theta_\zeta}{2} ,
\eea
and
\bea
\left< S^{z}_{\zeta} \right>_0 &=&\frac{\textrm{cos}\theta_\zeta}{2} .
\eea
Now, to obtain the ground-state phase diagram we calculate the ground-state 
energy to be 
\bea
E_g &=& \frac{\left<H\right>_0}{J^{'}_{1} N} \nonumber \\
    &=& -\frac{1}{4}\textrm{sin}\theta_a\textrm{sin}\theta_b
        -\frac{J^{'}_{2}}{8 J^{'}_{1}} \textrm{sin}^2\theta_a  
        -\frac{J^{'}_{2}}{8 J^{'}_{1}} \textrm{sin}^2\theta_b + \nonumber \\
    & & \frac{J^{'}_{z}}{8 J^{'}_{1}} \textrm{cos}\theta_a\textrm{cos}\theta_b
        -\frac{B}{2 J^{'}_{1}} \left[\textrm{cos}\theta_a + \textrm{cos}\theta_b 
+ 2 \right] ,
\label{groundeng}
\eea
where $J^{'}_{1} \equiv Z_{nn} J_1 $, $J^{'}_{2} \equiv Z_{nnn} J_2$, and
 $J^{'}_{z} \equiv Z_{nn} J_z$, with
$Z_{nn}$ and $Z_{nnn}$ being the number of nearest-neighbor(NN)  and 
next-nearest-neighbor (NNN) hopping processes respectively. For the Hamiltonian
 of Eq.(\ref{heff}), $Z_{nn}=4$ while
$Z_{nnn}=12$ because of the diagonal hoppings given by the third term
on the right-hand-side of 
Eq.(\ref{heff}).
Minimization of $E_g$, with respect to $\theta_a$ and $\theta_b$,
 gives the following two conditions:
\bea
2B \textrm{sin}\theta_a &=& J^{'}_{1}\textrm{cos}\theta_a\textrm{sin}\theta_b
+ J^{'}_{2}\textrm{sin}\theta_a \textrm{cos}\theta_a 
+ 0.5 J^{'}_{z}\textrm{sin}\theta_a \textrm{cos}\theta_b \nonumber \\
2B \textrm{sin}\theta_b &=& J^{'}_{1}\textrm{cos}\theta_b\textrm{sin}\theta_a
+ J^{'}_{2}\textrm{sin}\theta_b \textrm{cos}\theta_b 
+ 0.5 J^{'}_{z}\textrm{sin}\theta_b \textrm{cos}\theta_a .
\nonumber \\
& & 
\label{minimization}
\eea
Now we note that, for a charge density wave (CDW)
state we have $(\theta_a,\theta_b)=(0,\pi)$ or
 $(\theta_a,\theta_b)=(\pi,0)$; for a superfluid (SF) state  
 $\theta_a = \theta_b$; and
 for a phase separated (PS) regime  $\theta_a \neq \theta_b$ and $\theta_{a,b}
\neq 0$ or $\pi$.
Then, from Eq.(\ref{minimization}), we obtain the following expression
for the phase boundary: 
\bea
\frac{J^{'}_{z}}{2 J^{'}_{1}} - \frac{J^{'}_{2}}{J^{'}_{1}} 
=  \frac{J_{z}}{2 J_{1}} - \frac{3 J_{2}}{J_{1}} =
\frac{1+\left(2 n-1\right)^2}{1-\left(2 n-1\right)^2}  .
\label{phaseboundary}
\eea
From Eq. (\ref{phaseboundary}), we see that 
 we obtain
the mean-field phase boundary of Ref. [\onlinecite{micnas}]
when $J^{'}_{2}=0$. We further note that
NNN hopping does not change the qualitative feature of the 
\begin{figure}[]
\includegraphics[angle=0,width=0.8\linewidth]{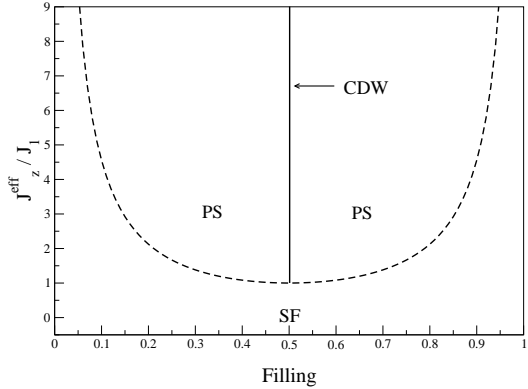}
\caption{Mean field phase diagram with
$J^{eff}_{z} = J_z/2 - 3 J_2$.}
\label{mfa_phase_diag.eps}
\end{figure}
phase diagram (see Fig. \ref{mfa_phase_diag.eps}); it
only increases the critical value of ${J^{}_{z}}/{J^{}_{1}}$ at 
which the transition from SF state to PS or CDW state occurs.
\section{\label{DLRO}Diagonal long range order and structure factor}
Diagonal long range order (DLRO) is the typical property of a crystalline solid and information about the
periodicity in the solid is contained in the structure factor.
For crystalline solids, the structure factor  shows delta function
peak at the reciprocal lattice points.
In terms of the particle density operators the structure factor is given by
\begin{equation}
S(\bf{q}) = 
\sum_{i,j} e^{\bf{q}\cdot(\bf{R_i} -\bf{R_j})}
(\left\langle n_i n_j\right\rangle -
\left\langle n_i\right\rangle \left\langle n_j\right\rangle ).
\label{sf}
\end{equation}
In this paper, we have calculated the structure factor using exact diagonalization technique 
for lattice clusters of size $4 \times 4$ , 
$\sqrt{18} \times \sqrt{18}$, and $\sqrt{20} \times \sqrt{20}$ . 
\subsection{\label{DLRO_j20}$J_2 = 0$}
\begin{figure}[]
\includegraphics[angle=0,width=0.8\linewidth]{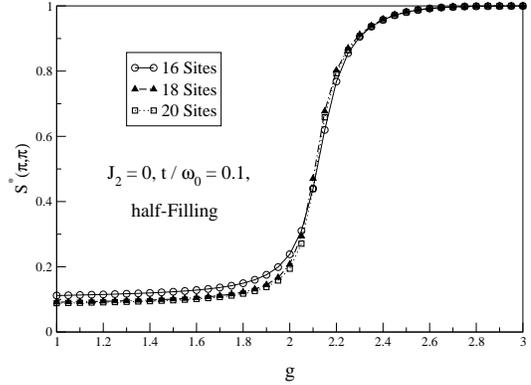}
\caption{Normalized structure factor
$S^{*}(\pi,\pi) = S(\pi,\pi)/S^{max}(\pi,\pi)$
for three different lattice clusters at half-filling. 
Here, adiabaticity parameter
$t/\omega_0=0.1$ and NNN hopping $J_2 = 0$.}
\label{sf_compare_J20_ad0.1}
\end{figure}

We will now present the structure factor results
 when $J_2 = 0$, i.e., for the xxz-model. 
In Fig. \ref{sf_compare_J20_ad0.1}, we have plotted the normalized structure factor 
$S^{*}(\pi,\pi) = S(\pi,\pi)/S^{max}(\pi,\pi)$ where 
$S^{max}(\pi,\pi)$ corresponds to all particles in only one sub-lattice. 
The calculations were done
 at half-filling
 and for different lattice clusters with the adabiticity parameter $t/\omega_0 = 0.1 $.
 From Fig. \ref{sf_compare_J20_ad0.1}, we see that, at half-filling, 
the system makes a transition 
to a CDW state at a critical b-p coupling strength  $g_c \approx 2.15  $. 
\begin{figure}[]
\includegraphics*[angle=0,width=0.8\linewidth]{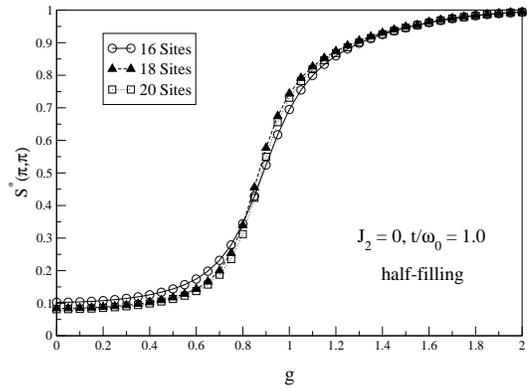}
\caption{Depiction of $S^{*}(\pi,\pi)$ for three different lattice clusters
 at half-filling with 
$t/\omega_0=1.0$ and $J_2 = 0$ .}
\label{sf_compare_J20_ad1.0}
\end{figure}
On the other hand, at a larger $t/\omega_0 = 1.0$, the transition for a
 half-filled system occurs at a significantly lower value of $g_c \approx 0.9$
(see Fig. \ref{sf_compare_J20_ad1.0}).
 This is because, for $J_2 = 0$, the
 transition is governed only by the ratio
 $J_z/J_1$.
Since $(J_z/J_1)\times (\omega_0/t)$ is a monotonically increasing function of
$g$, for a larger value of $t/\omega_0$, it takes a lower value of $g$
 to attain the same value of $J_z/J_1$.
 Another important point to note from Figs. \ref{sf_compare_J20_ad0.1}
 and \ref{sf_compare_J20_ad1.0} is that 
$S^{*}(\pi,\pi)$ is almost identical for different lattice clusters. The jump
 in the structure factor becomes sharper as we increase the system size.
 However this does not change the point of transition significantly.
 This means that the $4 \times 4$ 
lattice cluster is enough to have a reasonable estimate of the transition point.

\begin{figure}[]
\includegraphics*[angle=0,width=0.8\linewidth]{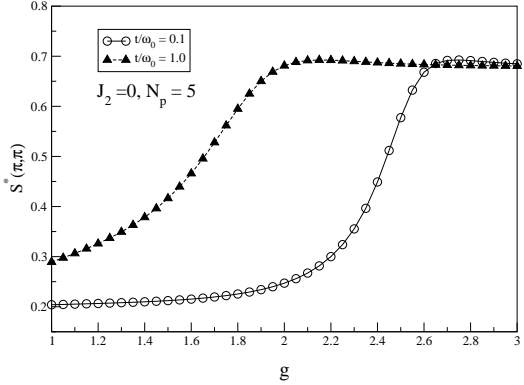}
\caption{Comparison of $S^{*}(\pi,\pi)$ values
for $t/\omega_0=0.1 \& 1.0 $. 
Plots are for $5$ particles in a
 $4 \times 4$ lattice  
 and $J_2 = 0$ .}
\label{sf_compare_J20_np5}
\end{figure}

 Next, we proceed to analyze the
 system away from half-filling. 
Without actually presenting the details of the calculations,
we first note
 that, for $N_p \le 4$
 in a $4 \times 4$ lattice, there is no evidence
of a phase transition.
 Here, we present the results for total number
of particles $N_p = 5$ in a $4 \times 4$ cluster.
 From Fig. \ref{sf_compare_J20_np5}, we see that
the qualitative features of the transition are similar
to those of the half-filled case.
However, in detail, the two cases
differ in the following sense.
Firstly, the critical values of the b-p coupling are larger for 
 $N_p =5$ with 
  $g_c \approx 2.45$ for $t/\omega_0 = 0.1$, while
 $g_c \approx 1.70$ for $t/\omega_0 =1.0$. 
Secondly, the $5$-particle system
 never attains
 a fully CDW-state as seen from  
$S^{*}(\pi,\pi)$ being noticeably less than unity.
Lastly, in  
  Fig. \ref{sf_compare_J20_np5}, we see that
$S^{*}(\pi,\pi)$ decreases 
 slowly after attaining a peak value which is perhaps due to
some special correlations which need to be examined.
\subsection{\label{DLRO_j2nonzero}$J_2 \neq 0$ }
\begin{figure}[]
\includegraphics*[angle=0,width=0.8\linewidth]{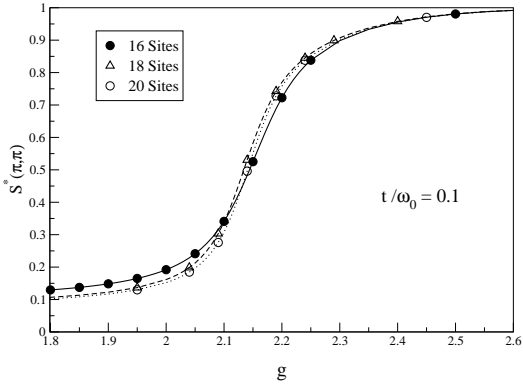}
\caption{$S^{*}(\pi,\pi)$ for three different lattice clusters at half-filling,
extreme adiabaticity ($t/\omega_0=0.1$), and non-zero NNN hopping
$J_2$ .}
\label{sf_compare_ad0.1}
\end{figure}
In this sub-section, we shall consider the effect of the
additional NNN hopping $J_2$.
At half filling, for a small value of the
 adabiticity parameter $t/\omega_0=0.1$, we find that
the system undergoes a phase transition from a SF-state to a
 CDW-state at
a critical boson-phonon coupling strength $g_c \approx 2.16$
(see Fig. \ref{sf_compare_ad0.1})
 which is 
very close to the case when $J_2=0$
  (see Fig. \ref{sf_compare_J20_ad0.1}).
 This is because, when $t/\omega_0$ is small, the ratio $J_2/J_1 << 1$ for
all values of $g$ (see Fig. \ref{constants}). 

\begin{figure}[]
\includegraphics*[angle=0,width=0.8\linewidth]{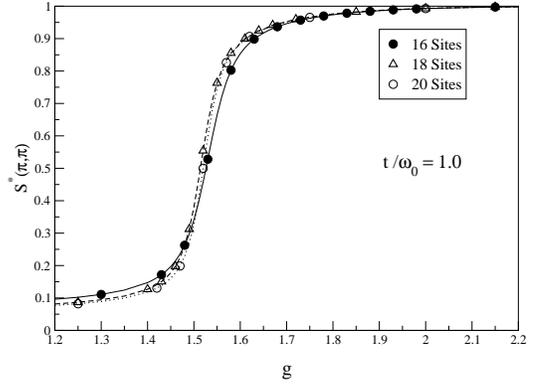}
\caption{Plots of $S^{*}(\pi,\pi)$ for three different lattice clusters at
 half-filling,  
$t/\omega_0=1.0$, and $J_2 \neq 0$ .}
\label{sf_compare_ad1.0}
\end{figure}

 When we increase the value of $t/\omega_0$,
 as is evident from Fig. \ref{sf_compare_ad1.0},
 the value of the
critical coupling $g_c$ decreases.
 The physical reason for this has already been discussed
 in section \ref{DLRO_j20}. 
 At half-filling, 
for $t/\omega_0=1.0$, we find $g_c \approx 1.53$.
\begin{figure}[]
\includegraphics*[angle=0,width=0.8\linewidth]{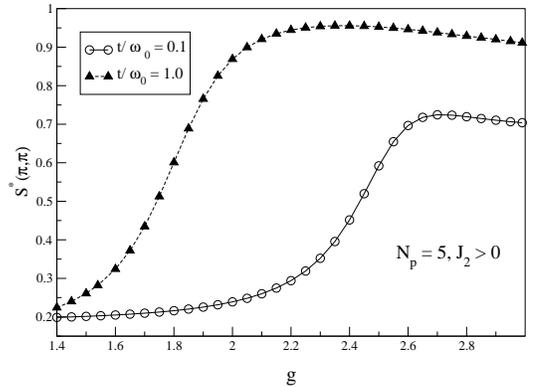}
\caption{Comparison of $S^{*}(\pi,\pi)$ plots
 at two extreme values of $t/\omega_0$.
Figures are for $5$ particles in a
$4 \times 4$ lattice cluster 
and $J_2 \neq 0$ .}
\label{sf_compare_np5}
\end{figure}
Furthermore, for a given value of $t/\omega_0$, the system makes a transition
 to the CDW state at  a lower value of $g$ when NNN hopping 
 $J_2 = 0$. This is in accordance with the mean field analysis, which shows
 that the presence of $J_2$ delays the transition
 [see Eq. (\ref{phaseboundary})].
 This is because, for relevant values of $g$, when $t/\omega_0 = 1$, 
 the $J_2/J_1$ term  is small but not negligible. Presence of $J_2$ introduces
 disorder in the system. Hence, it takes a higher $J_z/J_1$ ratio (i.e.,
 a higher value of $g$) to make the system ordered. 
Similar to the $J_2 = 0$ case 
 in a $4 \times 4$ lattice,
we also find that for fillings up to 0.25
 there is no transition to a CDW state while
a CDW transition does occur 
 for $5$ hcb.
 For $N_p=5$, as seen from Fig. \ref{sf_compare_np5},
 $g_c \approx 2.45$ for $t/\omega_0 =0.1$ while $g_c \approx 1.85$
 for $t/\omega_0 = 1.0$.
\section{\label{ODLRO}Off-diagonal long range order}
The concept of off-diagonal long range order (ODLRO) was
 introduced by Penrose and Onsager \cite{penrose} 
to understand the nature of the order in superfluids. Bose-Einstein 
condensate is one example which shows ODLRO. Following 
Refs. [\onlinecite{mahan}] and [\onlinecite{huang}],
 we define the general one-particle density matrix as 
\begin{equation}
\tilde \rho (i,j) = \left< b^{\dagger}_i b_j \right> = 
\frac{1}{N} \sum_{\mathbf k,\mathbf q} e^{(\mathbf k \cdot \mathbf R_i - \mathbf q \cdot \mathbf R_j)}
\left< b^{\dagger}_{\mathbf k} b_{\mathbf q} \right> .
\label{density_matrix}
\end{equation}
Here $\left< \right>$ denotes ensemble average and
 $b^{\dagger}_{\mathbf k}$ is the creation operator for hcb
 in momentum space.
It is easy to see that $\tilde \rho$ becomes the diagonal one-particle density matrix when 
$i = j$ . 
\subsection{\label{Condense_frac}Condensate fraction}
It follows from Eq. (\ref{density_matrix}) that, for a translationally 
invariant system,
\begin{equation}
\sum_j \tilde \rho (i,j) = \left<n_0\right> , 
\label{bec_frac}
\end{equation}
where $n_0$ is the occupation number for the $\mathbf k = 0$ momentum state.
Eq. (\ref{density_matrix})
 gives the Bose-Einstein condensate fraction as 
\begin{equation}
n_b=\sum_{i,j} \frac{\tilde \rho (i,j)}{N N_p} .
\label{n_b}
\end{equation}
In general, to find $n_b$, one
constructs the generalized one-particle density matrix $\tilde \rho$ and then diagonalizes it to find out
the largest eigenvalue. This procedure alone does not tell us which momentum state corresponds to the largest
eigenvalue. To find out whether the $\mathbf k =0$ momentum state is macroscopically occupied or not,
 we proceed as follows. First let us see if
$ \left<n_0\right> $ 
 is one of the eigenvalues.
 For this consider the following single-particle generalized density matrix,
\begin{equation*}
\tilde \rho = \left(\begin{array}{cccc}
              \tilde \rho(1,1) & \tilde \rho(1,2) & \cdots  & \tilde \rho(1,N) \\
              \tilde \rho(2,1) & \tilde \rho(2,2) & \cdots  & \tilde \rho(2,N) \\
              \vdots           & \vdots           & \vdots  & \vdots           \\
              \tilde \rho(N,1) & \tilde \rho(N,2) & \cdots  & \tilde \rho(N,N) \\
                    \end{array} \right), 
\end{equation*}
which can be re-written as 
\begin{equation}
\tilde \rho = \left(\begin{array}{cccc}
 \sum_i\tilde \rho(i,1) & \sum_i \tilde \rho(i,2) & \cdots  & \sum_i\tilde\rho(i,N)
 \\
 \tilde \rho(2,1) & \tilde \rho(2,2) & \cdots  & \tilde \rho(2,N) \\
 \vdots           & \vdots           & \vdots  & \vdots           \\
 \tilde \rho(N,1) & \tilde \rho(N,2) & \cdots  & \tilde \rho(N,N) \\
             \end{array} \right). 
\label{matrix}
\end{equation}
From Eq. (\ref{matrix}), it is easy to see that, for a translationally invariant system,
$ \left<n_0\right> $ is indeed 
 an eigenvalue of the one-particle generalized density matrix. 
 We found in our calculations that, 
 for all the relevant regions of the various parameter spaces,
the value of  $\left< n_0 \right>$ obtained according to Eq. (\ref{bec_frac})
 and the
 highest eigenvalue of the density matrix coincide quite accurately.
\subsection{\label{Sf_fraction}Superfluid fraction}
To characterize a superfluid, another important quantity of interest is the 
 superfluid fraction $n_s$ . The order parameter for a superfluid is a complex number and it is taken
 to be 
$\left< b \right> = \sqrt{\left<n_0 \right>/N} e^{i\theta} $ where the lattice constant has been taken
to be unity.
Spatial variation in the phase $\theta$ will increase
the free energy (or simply the energy at $T = 0$) of the system.
We consider an imposed phase 
variation that is a linear function
 of the phase angle, i.e., we
take $\theta(x) = \theta_0 \frac{x}{L}$  where $L$ is the linear dimension of the 
system along the $x$-direction. For simplicity we have chosen the variation in $\theta$ to be 
only along the $x$-direction. With these considerations we can write the change in energy to be 
\begin{equation}
 E[\theta_0] - E[\theta=0]=
\frac{1}{2} m N_p n_s \left|\frac{\hbar}{m} \vec\nabla \theta(x) \right|^2 ,
\end{equation}
where 
$E[\theta_0]$
 corresponds to an imposed phase variation
 when $\theta_0 \neq 0$.
Here, it is important to note that $\theta_0$ should be small, 
because a larger $\theta_0$ can
 induce other excitations which can destroy the collective motion of the superfluid 
component (see Ref. [\onlinecite{Roth}] and the references therein for details). 
We then get  the superfluid fraction 
 to be
\begin{equation}
n_s = \left(\frac{N}{N_p t_{eff}}\right)
\frac{E[\theta_0] - E[\theta=0]}{\theta_0^2},
\label{sffrac_ours} 
\end{equation}
where
 $t_{eff} = \hbar^2/2 m$. For our Hamiltonian in Eq. (\ref{heff}),
 we find 
$t_{eff} = J_1 + 8 J_2$.
Now, to introduce the phase variation, we impose twisted boundary conditions on the many-particle wave function.
A twist in the boundary conditions is gauge-equivalent to modifying the hopping terms in the 
Hamiltonian of Eq. (\ref{heff}). With this modification, the effective Hamiltonian becomes
\bea
H_{\theta} &=& -g^2 \omega_0 \sum_j n_j - 
         J_1 \sum_{j,\delta} e^{ i {\theta} \hat x \cdot {(\mathbf R_j -\mathbf 
R_{j+\delta})}}b^{\dagger}_j b_{j+\delta} \nonumber \\
        & & - J_2 \sum_{j,\delta,\delta' \neq \delta}
e^{ i {\theta} \hat x \cdot {(\mathbf R_{j+\delta^{'}} -\mathbf 
R_{j+\delta})}}
             b^{\dagger}_{j+\delta'}b_{j+\delta} \nonumber \\
& &
         - 0.5 J_z \sum_{j,\delta} n_j(1- n_{j+\delta}) , 
\eea
where $ \hat x$ is a unit vector in the $x$-direction.

\section{Energy consideration}\label{maxwell}
In this section, we shall examine the possibility of phase separation for the system of hcb
coupled with optical phonons. To this end,
 we have calculated the free energy for different number
 of particles in a $4 \times 4$ lattice. 
In Sec. \ref{DLRO}, for both $J_2 = 0$ and $J_2 > 0$,
 we observed that the system is always a pure superfluid for $N_p \le 4$
 in a $4 \times 4$
 lattice and that it is either a pure CDW or a pure superfluid at half-filling.
 After plotting the free energy at different fillings, if it is found that the curve is convex 
at a given filling, then the system at that filling is said to be stable; whereas, if
 the curve is concave at that filling, then the system would be unstable against phase separation.
 This procedure of calculating free energy at various fillings to figure out the stability
 of a system is equivalent to the well-known \textit{Maxwell construction}.
\subsection{$J_2 = 0$}\label{maxwell_j20_0}
\begin{figure}[]
\includegraphics*[angle=-90,width=0.8\linewidth]{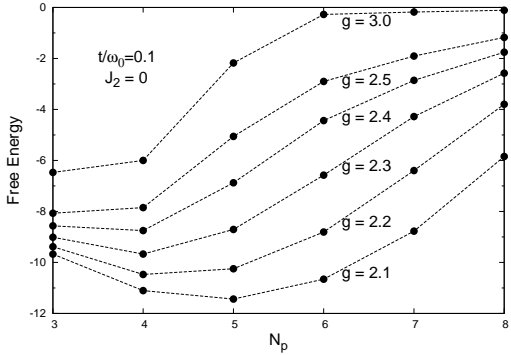}
\caption{Free energy plots at different fillings for $J_2 = 0$,
 $t/\omega_0 = 0.1$, and various b-p couplings $g$ . }
\label{psep_j20_ad0.1}
\end{figure}

\begin{figure}[]
\includegraphics*[angle=-90,width=0.8\linewidth]{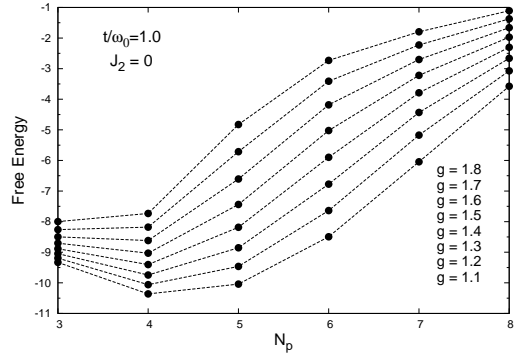}
\caption{Free energy versus particle number $N_p$
for parameters
 $J_2 = 0$, $t/\omega_0 = 1.0$, and $g$ at different values.
  }
\label{psep_j20_ad1.0}
\end{figure}

In Sec. \ref{DLRO_j20} we saw that, at half-filling, $J_2 = 0$,
 and $t/\omega_0 = 0.1$, the system
 makes a transition to a CDW state at $g \approx 2.15 $. 
From Fig. \ref{psep_j20_ad0.1},
we see that for $g \le 2.2$, the system in between quarter-filling and
 half-filling
 is stable due to the convexity of the energy curve here.
 As $g$ is increased, the system close to half-filling becomes unstable
first and then the 
lower-fillings
becoming unstable progressively.
For $J_2 = 0$, we see from Figs. \ref{psep_j20_ad0.1} and \ref{psep_j20_ad1.0} that the qualitative
 nature of the free energy curves, with respect to phase separation, does not depend on the adiabiticity
 parameter. 
 As expected from the explanation in Sec. \ref{DLRO}, 
 these figures show that the critical value $g_c$ (where the phase separation starts)
 decreases as $t/\omega_0$ increases. 
We notice in Fig. \ref{psep_j20_ad0.1}
(Fig. \ref{psep_j20_ad1.0}) 
that, for $N_p=7,6,\& 5$,
the energy curves become concave before $g=2.3,2.4, \& 2.5$ 
($g=1.2, 1.4, \& 1.7$) respectively.
Furthermore, for $N_p = 5$, the phase separation seems to occur at approximately the
same value of $g$ at which the CDW transition occurs (see Fig. \ref{sf_compare_J20_np5}).
\subsection{$J_2 \neq 0$}\label{maxwell_j2_nonzero}
\begin{figure}[]
\includegraphics*[angle=-90,width=0.8\linewidth]{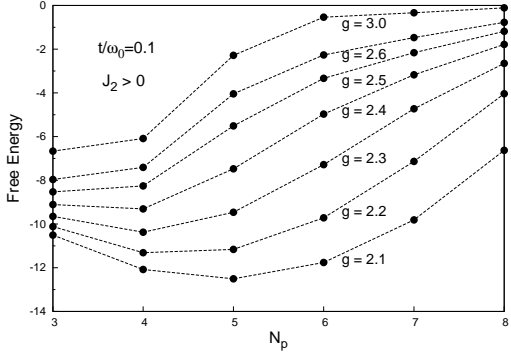}
\caption{Free energy at different fillings for
 $t/\omega_0 = 0.1$ and different values of $g$.
Here NNN hopping
 $J_2 \neq 0$.
  }
\label{psep_ad0.1}
\end{figure}

\begin{figure}[]
\includegraphics*[angle=-90,width=0.8\linewidth]{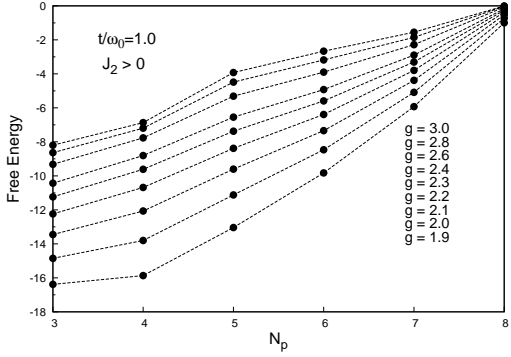}
\caption{Free energy versus filling
with $t/\omega_0 = 1.0$ and b-p coupling $g$ at different values.
 Here too  $J_2 > 0$. }
\label{psep_ad1.0}
\end{figure}

For the extreme anti-adiabatic regime,
 the situation,
when NNN hopping $J_2 \neq 0$,
 is not too different from the $J_2 = 0$ case.
This can be seen by comparing
Figs. \ref{psep_ad0.1}
and \ref{psep_j20_ad0.1} drawn for
 $t/\omega_0 = 0.1$.
This is expected because,
for the extreme anti-adiabatic regime, the ratio $J_2/J_1 << 1$.
However, when $t/\omega_0 = 1.0 $ ($J_2/J_1$ ratio is not negligible at values of $g$
considered in Fig. \ref{psep_ad1.0}),
 the situation is quite
different from the $J_2 = 0$ case away from half-filling. 
In Fig. \ref{sf_compare_np5}, we saw that
the structure factor revealed a CDW
transition at  
 $g_c \approx 1.85$ for $N_p = 5$.
However, for $J_2 \neq 0$, the phase separation occurs at a higher value of $g \approx 2.1$
as can be seen 
from Fig. \ref{psep_ad1.0}.
The interesting implications of the CDW transition occurring before the PS transition will be discussed in the
next section.

 One additional important feature for the $J_2 \neq 0$ case,
 compared to the $J_2 = 0 $ case, is that
 the 
 phase separation first occurs
 at the low-filling side. For $6$ particles, 
the phase separation sets in at $g \approx 3.0$; 
while for $7$ particles the corresponding
$g$-value is expected to be even higher. We could not obtain the exact $g$-value for PS instability for $N_p=7$
as our code had convergence problems 
 when we tried to go beyond $g=3.0$. 
\section{Results and discussions}\label{Results}
\begin{figure}[]
\includegraphics*[angle=0,width=0.8\linewidth]{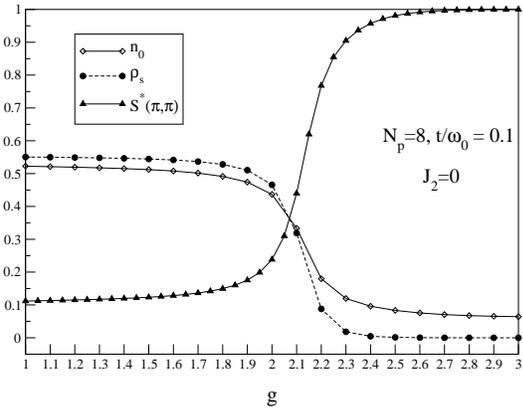}
\caption{Comparison of normalized structure factor
$S^{*}(\pi,\pi)$, condensate fraction $n_b$,  and superfluid fraction
$ n_s$ for $8$ particles
when $J_2 = 0$ and $t/\omega_0 = 0.1.$}
\label{np8_bec.rho.sf_j20_ad0.1.eps}
\end{figure}
\begin{figure}[]
\includegraphics*[angle=0,width=0.8\linewidth]{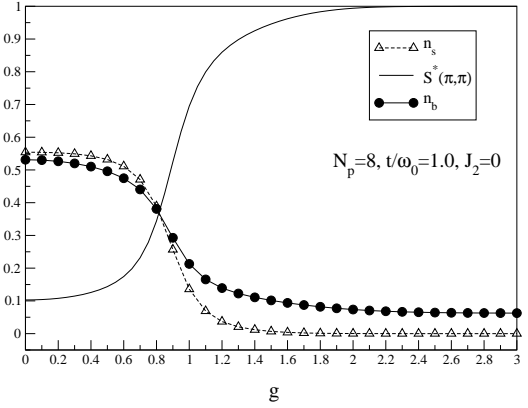}
\caption{Comparison of $S^{*}(\pi,\pi)$, $n_b$, and $n_s$ for $8$ particles
when $J_2 = 0$ and $t/\omega_0 = 1.0.$}
\label{np8_bec.rho.sf_ad1.0_j20.eps}
\end{figure}
\begin{figure}[]
\includegraphics*[angle=0,width=0.8\linewidth]{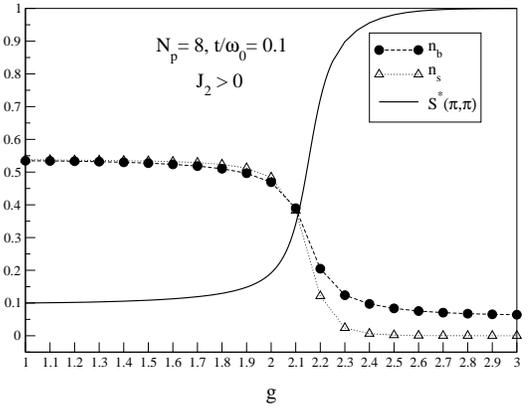}
\caption{Comparative plots of $S^{*}(\pi,\pi)$, $n_b$,
 and $n_s$ for $8$ particles
when $t/\omega_0 = 0.1$, but
$J_2 > 0$.}
\label{np8_bec.rho.sf_ad0.1.eps}
\end{figure}
\begin{figure}[]
\includegraphics*[angle=0,width=0.8\linewidth]{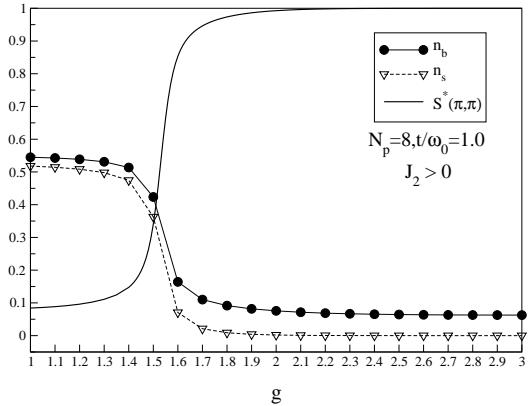}
\caption{Comparative depiction of $S^{*}(\pi,\pi)$, $n_b$,
  and $ n_s$ for $8$ particles
when $t/\omega_0 = 1.0$
and $J_2 > 0$.}
\label{np8_bec.rho.sf_ad1.0.eps}
\end{figure}
\begin{figure}[]
\includegraphics*[angle=0,width=0.8\linewidth]{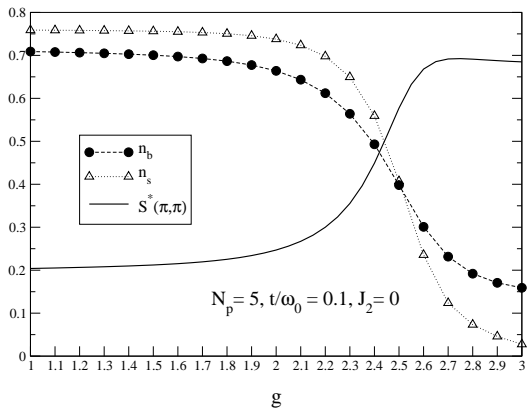}
\caption{Concomitant transitions depicted by
$S^{*}(\pi,\pi)$, $n_b$, and $n_s$ for $5$ particles
when $J_2 = 0$ and $t/\omega_0 = 0.1.$}
\label{np5_bec.rho.sf_ad0.1_j20.eps}
\end{figure}
\begin{figure}[]
\includegraphics*[angle=0,width=0.8\linewidth]{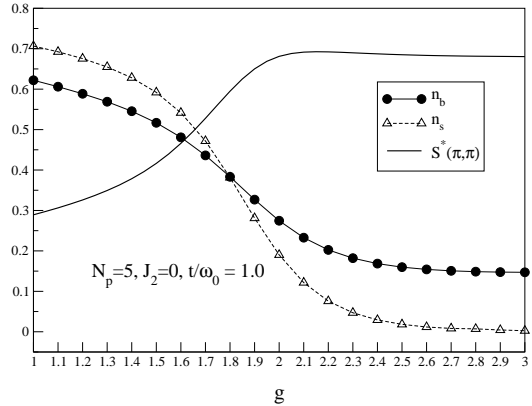}
\caption{Simultaneous phase transitions shown by
 $S^{*}(\pi,\pi)$, $n_b$, and $n_s$ for $5$ particles
when $J_2 = 0$ and $t/\omega_0 = 1.0.$}
\label{np5_bec.rho.sf_j20_ad1.0.eps}
\end{figure}
\begin{figure}[]
\includegraphics*[angle=0,width=0.8\linewidth]{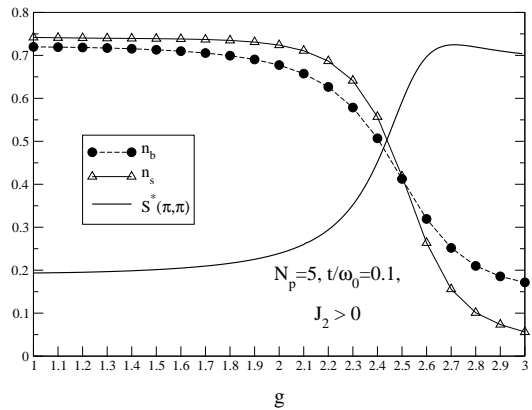}
\caption{Comparative study of $S^{*}(\pi,\pi)$, $n_b$, and $n_s$ for $5$
 particles
when $J_2 \neq 0$ and $t/\omega_0 = 0.1.$}
\label{np5_bec.rho.sf_ad0.1.eps}
\end{figure}
\begin{figure}[]
\includegraphics*[angle=0,width=0.8\linewidth]{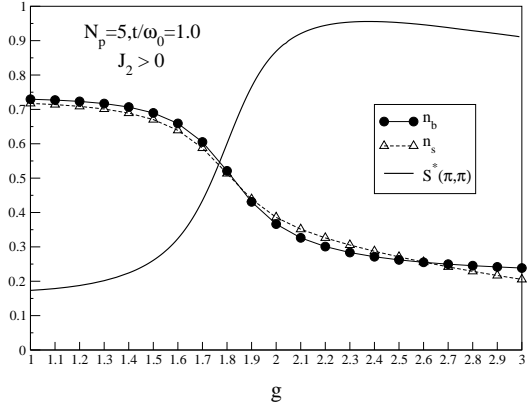}
\caption{Comparison of $S^{*}(\pi,\pi)$, $n_b$, and $n_s$ for $5$ particles
when $J_2 \neq 0$ and $t/\omega_0 = 1.0.$}
\label{np5_bec.rho.sf_ad1.0.eps}
\end{figure}
\begin{figure}[]
\includegraphics*[angle=0,width=0.8\linewidth]{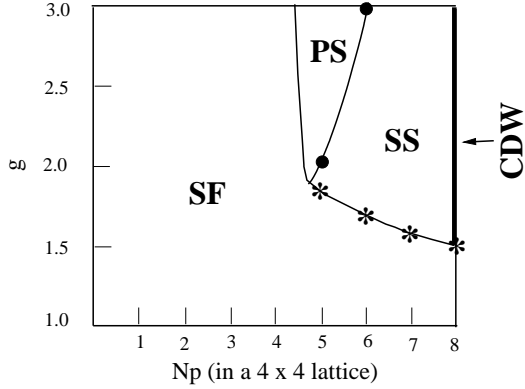}
\caption{Phase diagram depicting various phases for $J_2 > 0$ and $t/\omega_0 = 1.0$
 }
\label{phase-diag2}
\end{figure} 
\begin{figure}[]
\includegraphics*[angle=0,width=0.8\linewidth]{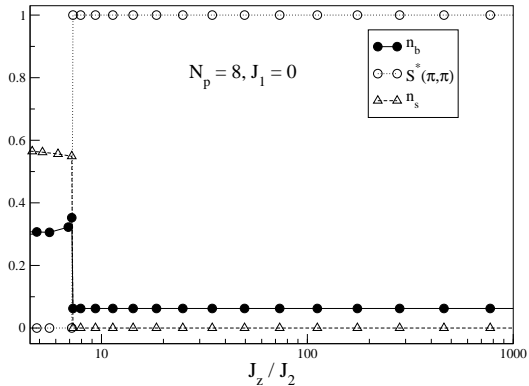}
\caption{Comparison of $S^{*}(\pi,\pi)$, $n_b$, and $n_s$ for $8$ particles
when $t/\omega_0 = 1.0$, but
 $J_1 = 0$. }
\label{np8_nbns_j10_ad1.0.eps}
\end{figure} 
\begin{figure}[]
\includegraphics[angle=0,width=0.8\linewidth]{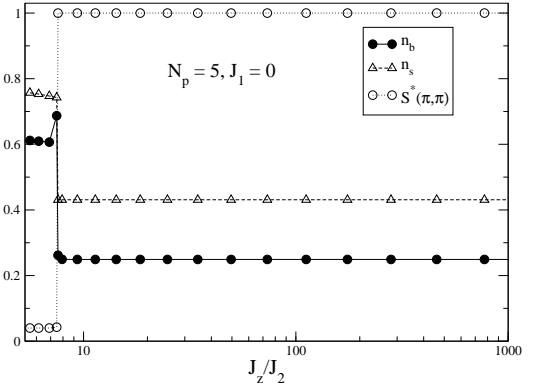}
\caption{Comparison of $S^{*}(\pi,\pi)$, $n_b$, and $ n_s$ for $5$ particles
when $t/\omega_0 = 1.0$.
Here too $J_1 = 0$.}
\label{np5_nbns_j10_ad1.0.eps}
\end{figure}
\begin{figure}[]
\includegraphics[angle=-90,width=0.8\linewidth]{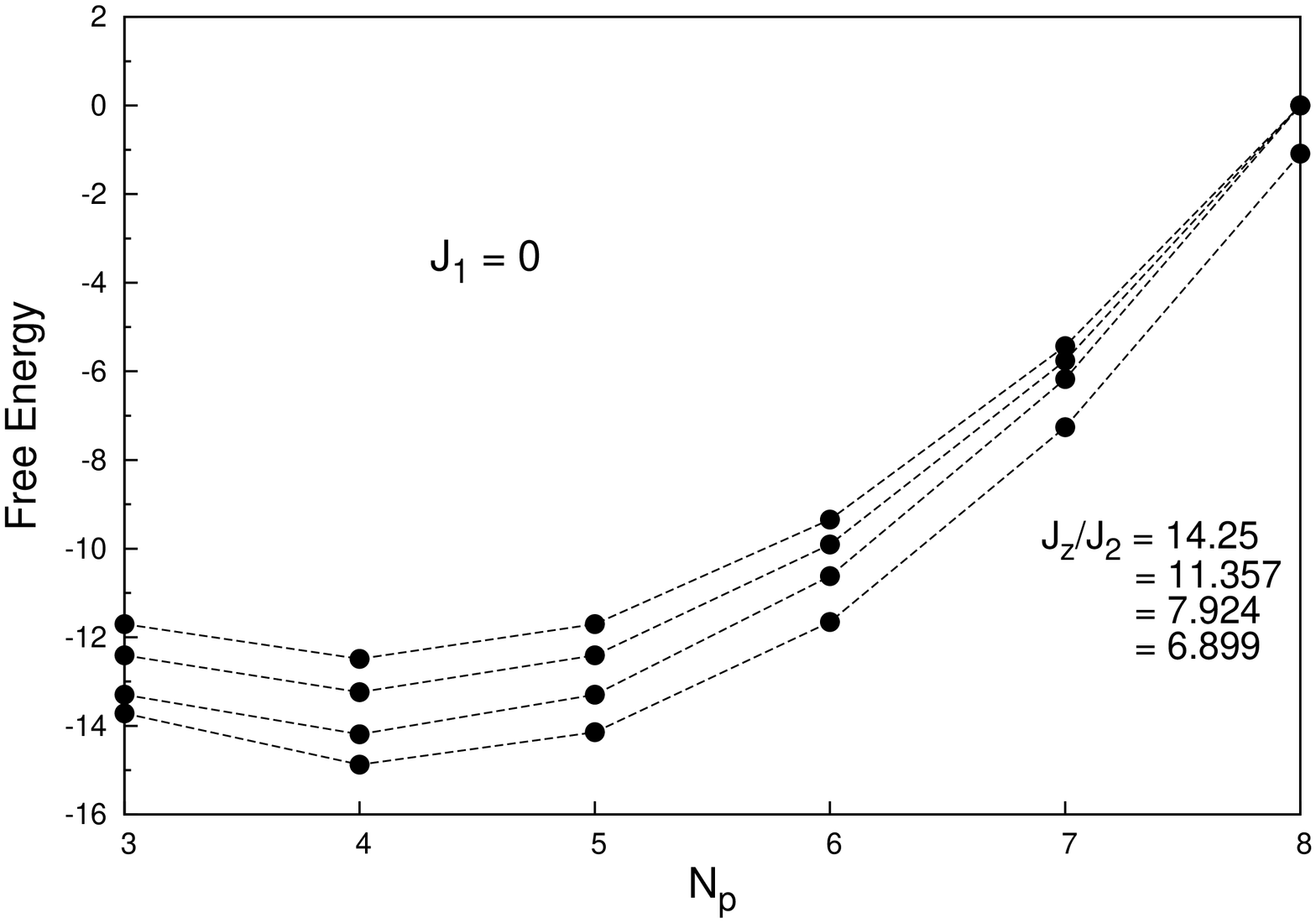}
\caption{Plot of free energy for different number of particles at various
values of $g$
 when $J_1 = 0.$}
\label{psep_j10_ad1.0.eps}
\end{figure}
Here, we will analyze together, in one plot, the quantities
 $S^{*}(\pi,\pi)$, $n_b$ and $n_s$ that were presented in earlier sections. 
For a half-filled system at $J_2 = 0$
and $t/\omega_0 = 0.1$ ($t/\omega_0 = 1.0$), 
 we can see from Fig. \ref{np8_bec.rho.sf_j20_ad0.1.eps} 
(Fig. \ref{np8_bec.rho.sf_ad1.0_j20.eps})
that
the system undergoes a sharp transition to an
 insulating CDW state at
 $g_c \approx 2.15$ ($g_c \approx 0.9$). At $g=g_c$, while there is a sharp rise in
the structure factor $S(\pi,\pi)$, there is also a concomitant sharp drop
in both the condensation 
fraction $n_b$ and the superfluid fraction $n_s$.
Furthermore, while $n_s$ actually goes to zero, $n_b$ remains finite
[as follows from Eq. (\ref{n_b})] 
 at a value $1/N = 1/16$ which is
 an artifact of the finiteness of the system.
 The lower critical value of $g$  at higher values of the adiabaticity parameter
$t/\omega_0$ has already been explained in Sec. \ref{DLRO}. 
Thus at half-filling,
in the absence of NNN hopping, a system of hard core bosons coupled with optical phonons
 undergoes a transition from a superfluid state to an insulating CDW state.
 For a half-filled system,
 the presence of NNN hopping does not produce a qualitative 
difference in the plots,
 except for changing the critical value $g_c$ of transition
 and that too only for adiabaticity values $t/\omega_0$ of the order of unity.
 This can be seen from Figs. \ref{np8_bec.rho.sf_ad0.1.eps} and
 \ref{np8_bec.rho.sf_ad1.0.eps}.
 Here also $n_b$ and $n_s$ behavior complements that of $S(\pi,\pi)$; the values of $n_b$ and $n_s$
drop noticeably when $S(\pi,\pi)$ increases sharply.
 These results for half-filling, with $J_2 = 0$ and $J_2 \neq 0$, were already qualitatively
 predicted in the mean-filed analysis of Sec. \ref{MFA}.

 Away from half-filling, the system shows markedly 
different behavior compared to the half-filled
situation.
 From Figs. \ref{np5_bec.rho.sf_ad0.1_j20.eps} 
and \ref{np5_bec.rho.sf_j20_ad1.0.eps},
 for $J_2 = 0 $, although $S(\pi,\pi)$ displays a CDW transition at
a critical value $g_c$, $n_b$ does not go to zero,
 again due to finite size effects,
 even at large values of $g$.
For $N_p = 5$ and $t/\omega_0 = 0.1$ ($t/\omega_0 = 1.0$),  we obtain
the critical value 
$g_c = 2.45$ ($g_c = 1.75)$. 
From Figs. \ref{psep_j20_ad0.1} and \ref{psep_j20_ad1.0}, we see clearly that
at these critical values of $g$, the free energy curves become concave for $N_p = 5$.
This suggests that 
the system is in a phase-separated state, i.e.,
 it is an inhomogeneous mixture of CDW-state and superfluid-state.
 Thus away from half filling, when $J_2 = 0$, our hcb-system
 undergoes a transition from a superfluid-state to a phase-separated-state at a critical boson-phonon
 coupling strength.

 In the presence of NNN hopping and in the extreme anti-adiabatic 
limit also, the system's
behavior for $N_p = 5 $ is very similar to that of $J_2 =0$
 at the same adiabaticity
as can be seen by comparing
 Fig. \ref{np5_bec.rho.sf_ad0.1.eps} with 
Fig. \ref{np5_bec.rho.sf_ad0.1_j20.eps} 
 and Fig. \ref{psep_ad0.1} with Fig. \ref{psep_j20_ad0.1}.
 However for $t/\omega_0$ not too small, when NNN hopping is present, the system shows a 
strikingly new 
 behavior for a certain region of the $g$-parameter space.
 Let us consider the system at $N_p=5$,
 $t/\omega_0 = 1.0$, and $J_2 \neq 0$. 
 Fig. \ref{np5_bec.rho.sf_ad1.0.eps}
shows that,
 above $g \approx 1.85$, the system enters a CDW state
 (as can be seen from the structure factor); however, it
continues to have a superfluid character as reflected by
the finite value of $n_s$.
Furthermore, Fig. \ref{psep_ad1.0} reveals that the 
system is phase-separated only above $g = 2.0$.
This simultaneous presence of DLRO and ODLRO, without any inhomogeneity (for $1.85 < g < 2.1$),
  implies that the system is a
{\em  supersolid}. 
 Similarly, for $6$ and $7$ particles as well,
we find that the system undergoes transition from a superfluid- to a supersolid-state and then to
a phase-separated-state. This is displayed in the phase diagram given in Fig. \ref{phase-diag2}.

 Another point to be noted here is that,
 in Figs.  \ref{np8_bec.rho.sf_ad1.0.eps} and \ref{np5_bec.rho.sf_ad1.0.eps}
(i.e., for $t/\omega_0 =1.0$, $J_2 > 0$, and
 small values of $g$), $n_s$ becomes smaller
 than $n_b$. We feel that this is an artifact of the approximation
 used for the mass
 in Eq. (\ref{sffrac_ours}). 

Finally, we shall present the interesting 
case of $J_1 =0$ as a means of understanding the supersolid phase
in the phase diagram of Fig. \ref{phase-diag2}.
 The physical scenario,
 when $J_1$ can be negligibly small compared
 to $J_2$ (see Ref. [\onlinecite{sudhakar1}] for a one-dimensional example), 
 and the detailed results 
will be published later \cite{sdyspbl}. 
Here we only present the results that are
relevant to the conclusions 
 made in the above discussions.
It is quite natural that, when $J_1 = 0$, all the particles will occupy a single sub-lattice 
for large 
values of nearest-neighbor repulsion.
 For a half-filled system, above a critical point,
 all the particles get localized, resulting in an insulating state. 
This can be seen from Fig. \ref{np8_nbns_j10_ad1.0.eps}.
 One can see that (at $J_z/J_2 \approx 7.2$)
 the structure factor dramatically jumps to its maximum value,
 while $n_s$ drops to zero and
  $n_b$ takes
 the limiting value of $1/16$ for 
reasons discussed earlier. This shows that above 
 $J_z/J_2 = 7.2$,
 the system is in a insulating state 
with one sub-lattice being completely full. 
However, away from half-filling, the system
 conducts perfectly while occupying a single sub-lattice
because of the presence of holes in the sub-lattice.
For instance, from Fig. \ref{np5_nbns_j10_ad1.0.eps} 
drawn for $N_p=5$, we see that the structure 
factor jumps to its maximum value at $J_z/J_2 \approx 7.5$, while $n_s$ 
drops
to a finite value which remains constant above $J_z/J_2 = 7.5$.
 From Fig. \ref{psep_j10_ad1.0.eps} we see, based on the curvature of the free energy curves,
 that the 5-particle system does not phase separate both above and below the transition.
 In fact, this single-phase-stability is true for any filling.
 This means that, at non-half filling and above a critical $J_z/J_2$, the system is homogeneous with
 simultaneous existence of both DLRO and ODLRO, i.e., the system exhibits supersolidity!
 Thus, except for the pathological case of one particle, the system
 undergoes a first-order phase transition from a superfluid to a supersolid state away
 from half-filling.
\section{ACKNOWLEDGMENTS}
S. Datta would like to thank Arnab Das for very useful 
discussions regarding numerical implementation of exact diagonalization.
S. Yarlagadda thanks K. Sengupta and S. Sinha for valuable discussions.


\end{document}